# Manifestation of vortex depinning transition in nonlinear current-voltage characteristics of polycrystalline superconductor Y$_{1-x}$Pr$_x$Ba$_2$Cu$_3$O$_{7-\delta}$

**V. A. G. Rivera[1], C. Stari[1,2], S. Sergeenkov[3,*], E. Marega[4], and F. M. Araújo-Moreira[1]**

[1]*Grupo de Materiais e Dispositivos, Departamento de Física, UFSCar, Caixa Postal 676, 13565-905, São Carlos, SP, Brasil*

[2]*Instituto de Física, Facultad de Ingeniería, Julio Herrera y Reissig 565, C.C. 30, 11000, Montevideo, Uruguay*

[3]*Departamento de Física, CCEN, Universidade Federal da Paraíba, Cidade Universitária, 58051-970 João Pessoa, PB, Brasil*

[4]*Instituto de Física, U SP, Caixa Postal 369, 13560-970, São Carlos, SP, Brasil*

**Abstract**

We present our recent results on the temperature dependence of current-voltage characteristics for polycrystalline Y$_{1-x}$Pr$_x$Ba$_2$Cu$_3$O$_{7-\delta}$ superconductors with $x = 0.0$, $0.1$ and $0.3$. The experimental results are found to be reasonably well fitted for all samples by a power like law of the form $V = R(I - I_c)^{a(T)}$. Here, we assume that $a(T)=1+ \Phi_0 I_C(T)/2\pi k_B T$ and $I_C(T)=I_C(0)(1-T/T_C)^{3/2}$ for the temperature dependences of the power exponent and critical current, respectively. According to the theoretical interpretation of the obtained results, nonlinear deviation of our current-voltage characteristics curves from Ohmic behavior (with $a(T_C)=1$) below $T_C$ is attributed to the manifestation of dissipation processes. They have a characteristic temperature $T_p$ defined via the power exponent as $a(T_p)=2$ and are related to the current induced depinning of Abrikosov vortices. Both $T_C(x)$ and $T_p(x)$ are found to decrease with an increase of Pr concentration $x$ reflecting deterioration of the superconducting properties of the doped samples.

**Keywords:** Y$_{1-x}$Pr$_x$Ba$_2$Cu$_3$O$_{7-\delta}$; Current-voltage characteristics; Vortex depinning

**PACS classification codes**: 74.72.Hs; 74.78.Bz; 74.40.+k

---
* Corresponding author: phone: (83) 3216-7544 ; fax: (83) 3216-7542; E-mail: sergei@fisica.ufpb.br

1. Introduction

The partial substitution of Y for Pr in the classical cuprate superconductor YBa$_2$Cu$_3$O$_{7-\delta}$ (YBCO) is usually used to shed some light on the pairing mechanism in the hole-dominated superconductivity of YBCO. It allows to monitor Pr induced destruction of hole-like Cooper pairs (within CuO plane) in the electron-doped superconductor Y$_{1-x}$Pr$_x$Ba$_2$Cu$_3$O$_{7-\delta}$ (for recent discussion of this problem, see, e.g., [1,2] and further references therein).

On the other hand, it is of great importance (from both fundamental and application points of view) to study the evolution of the different current transfer driven dissipation processes in doped superconductors [3]. Probably one of the most recognized dissipation mechanisms is the so-called Kosterlitz-Thouless (KT) topological transition. It is related to creation (destruction) of bound vortex-antivortex pairs below (above) some temperature $T_{KT}$ [4,5]. This transition usually manifests itself in sufficiently thin films via nonlinear current-voltage characteristics of the form $V = RI^{a(T)}$ with the power exponent markedly jumping from $a=1$ at $T=T_C$ to $a=3$ at $T=T_{KT}$. It should be noted, however, that in real materials (even in ultrathin films) the conventional KT transition can be easily masked by other mechanisms, such as interlayer coupling, size effects, extrinsic and intrinsic weak links, thermal fluctuations, quasi-particle contributions, etc. [3, 6-9]. For example, Repaci *et al.* [6] attributed the absence of the KT transition in their perfect ultrathin films to the existence of some competitive mechanisms (related to thermally activated motion of free vortices). At the same time, Medvedyeva *et al.* [7] argue that the results of Repaci *et. al.* [6] were strongly influenced by substantial finite-size effects in their ultrathin films which thus precluded them to observe the conventional KT transition.

In this paper we study the influence of Pr on hole dominated YBCO superconductor through measurements of current-voltage characteristics of the doped polycrystalline Y$_{1-x}$Pr$_x$Ba$_2$Cu$_3$O$_{7-\delta}$ sample. The obtained experimental results and their theoretical interpretation strongly suggest a possible manifestation of 3D (rather than 2D) dissipation effects in our polycrystalline samples related to current driven thermally activated motion of Abrikosov vortices.

## 2. Samples characterization and transport measurements

High quality $Y_{1-x}Pr_xBa_2Cu_3O_{7-\delta}$ bulk polycrystalline samples have been prepared by following a chemical route based on the polymeric precursors method (PECHINI). This method allows the attainment of more homogeneous samples (in comparison with other methods) through a more effective elimination of secondary phases. The phase purity and the structural characteristics of our samples were confirmed by both scanning electron microscopy (SEM) and x-ray diffraction (XRD). In this last case, we have also performed the standard Rietveld analysis. The analysis of the XRD data (Fig.1) shows that no secondary phases are present in our samples and that the peaks correspond to the orthorhombic structure with Y-123 stoichiometric phase. According to our results based on the resistivity measurements, the Pr substitution into Y chain sites of YBCO leads to quite a noticeable decrease of the bulk critical temperature $T_C(x)$ without inducing an orthorhombic–tetragonal phase transition with increasing of Pr (at least, up to $x=0.3$). More precisely, to determine the superconducting transition temperatures $T_C(x)$ for our samples, we measured the resistivity versus temperature by applying a small ac current (with amplitude of 2.5 mA and frequency of 20 Hz) by using the standard lock-in technique. The onset temperatures for all studied samples are listed in Table 1. Notice that they are well correlated with the values reported in the literature for polycrystalline samples with similar composition [2]. We have studied the current-voltage characteristics in a narrow temperature interval near the critical temperature $T_C(x)$, by using a sufficiently small dc current. The voltage signal was measured using a high-precision nanovoltmeter. Temperature was kept constant within a precision of ±0.1 K in all experiments.

## 3. Results and Discussion

Some typical results for current-voltage characteristics taken at different temperatures near $T_C(x)$ are shown in Fig. 2. As expected, in the normal state (above $T_C$) all the curves have an Ohmic behavior with $V = RI$, while below $T_C(x)$ the current-voltage characteristics show a nonlinear behavior for all samples with:

$$V = R(I - I_c)^{a(T)} \qquad (1)$$

Given the 3D nature of our polycrystalline samples, it is quite natural to assume that a power-like dependence of the current-voltage curves in the superconducting state is caused by a thermally activated depinning of Abrikosov vortices by applied currents (exceeding the critical currents $I_C(T)$ at a given temperature). This results in the critical current dependent power exponent. Namely, within this scenario, we postulate the following dependence [6]:

$$a(T) = 1 + \frac{U(T)}{k_B T} \qquad (2)$$

where

$$U(T) = \frac{\Phi_0 I_C(T)}{2\pi} \qquad (3)$$

is the current induced activation energy [10]. After trying many different forms for the temperature dependence of $I_C(T)$, we have found that all our data for the current-voltage curves are rather well fitted assuming the following dependence:

$$I_C(T) = I_C(0)\left(1 - \frac{T}{T_C}\right)^{3/2} \qquad (4)$$

which is also related to the well-known [10] Ginzburg-Landau (GL) expressions for the thermodynamic critical field $H_C(T) = H_C(0)(1 - T/T_C)$ and the in-plane penetration depth $\lambda_{ab}(T) = \lambda_{ab}(0)(1 - T/T_C)^{-1/2}$ as follows $I_C(T) = H_C(T)/\lambda_{ab}(T)$. Besides, within the GL theory, $H_C(T) = \Phi_0/2\pi\lambda_{ab}(T)\xi_{eff}(T)$, where the effective coherence length is given by $\xi_{eff}(T) = (\xi_{ab}\xi_c)^{1/2} = \xi_{eff}(0)(1 - T/T_C)^{-1/2}$. It accounts for the observed Pr induced *interlayer* weakening of the superconductivity in our doped samples. Recall that in undoped YBCO samples [11], $\xi_{ab}(0) = 2nm$ and $\xi_c(0) = 0.8nm$ while $\lambda_{ab}(0) = 0.14\mu m$. It is important to emphasize that the effective length scale is more appropriate for a realistic description of dissipation processes due to the depinning and free motion of 3D Abrikosov vortices. More precisely, within the transport current mediated scenario, the vortices start to penetrate the sample for $I > I_{C1}(T)$. In this case, $I_{C1}(T) = H_{C1}(T)/\lambda_{ab}(T)$ which is related to the Abrikosov lower critical field $H_{C1}(T) = \Phi_0/2\pi\lambda^2_{ab}(T)$. Some of the vortices will get pinned by defects through the pinning force $f_p(T)=\Phi_0 I_C(T)$ acting on each vortex. The depinning occurs when the current induced Lorentz force $f_L=\Phi_0 I$ overcomes the pinning force, i.e., for $I>I_C(T)$. In agreement with our observations, the above-suggested scenario results in the Ohmic behavior of the normal state with $I_C(T_C)=0$. Consequently, $a(T)$ starts to deviate from the linear $V(I)$ law for $T<T_C$. In particular, Fig.2 shows that the current-voltage dependence turns quadratic at some temperature $T_p$ which is responsible for the initiation of the vortex depinning processes. This is defined via the current induced power exponent as $a(T_p)=2$ or approximately by $T_p/T_C=1-[2\pi k_B T_C/\Phi_0 I_C(0)]^{2/3}$. Figure 3 shows the dependence of the normalized critical current $I_C(T)/I_C(0)$ and the power exponent $a(T)$ on reduced temperature $T/T_C$. Those values were obtained from the fittings of current-voltage curves shown in Fig.2 by using Eqs.(1)-(4). Notice that, in view of the above definition, the depinning temperature $T_p$ decreases (like $T_C$) while the difference $T_C-T_p$ increases with Pr doping, $x$. The values of $T_C$ extracted from the resistivity data along with the estimates for fitting and deduced parameters are summarized in Table 1. It is also worth mentioning that the doping induced evolution of the extracted parameters is in good agreement with those reported in the literature for similar compositions [1,2].

## 4. Conclusions

In summary, a possible manifestation of current induced depinning of Abrikosov vortices was observed in the temperature dependence of nonlinear current-voltage characteristics for polycrystalline $Y_{1-x}Pr_xBa_2Cu_3O_{7-\delta}$ superconductors with $x = 0.0$, $0.1$ and $0.3$. Both the superconducting transition $T_C(x)$ and depinning $T_p(x)$ temperatures were found to decrease with an increase of Pr concentration $x$ due to the expected weakening of the interlayer mediated superconducting properties of the doped samples.

**Acknowledgments**

This work has been financially supported by the Brazilian agencies CNPq, CAPES and FAPESP.

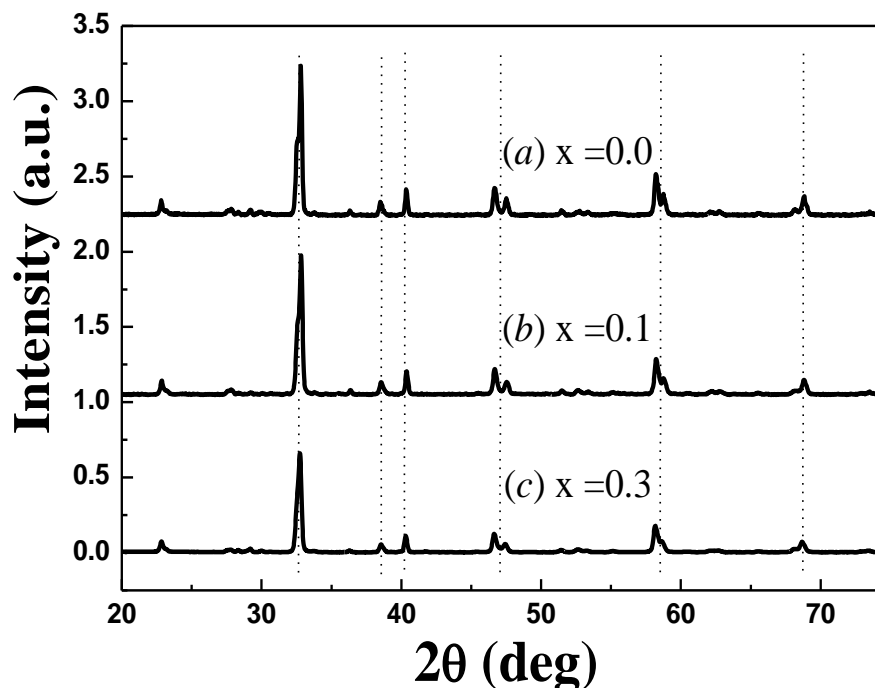

**Fig.1:** XRD patterns for $Y_{1-x}Pr_xBa_2Cu_3O_{7-\delta}$ polycrystalline samples.

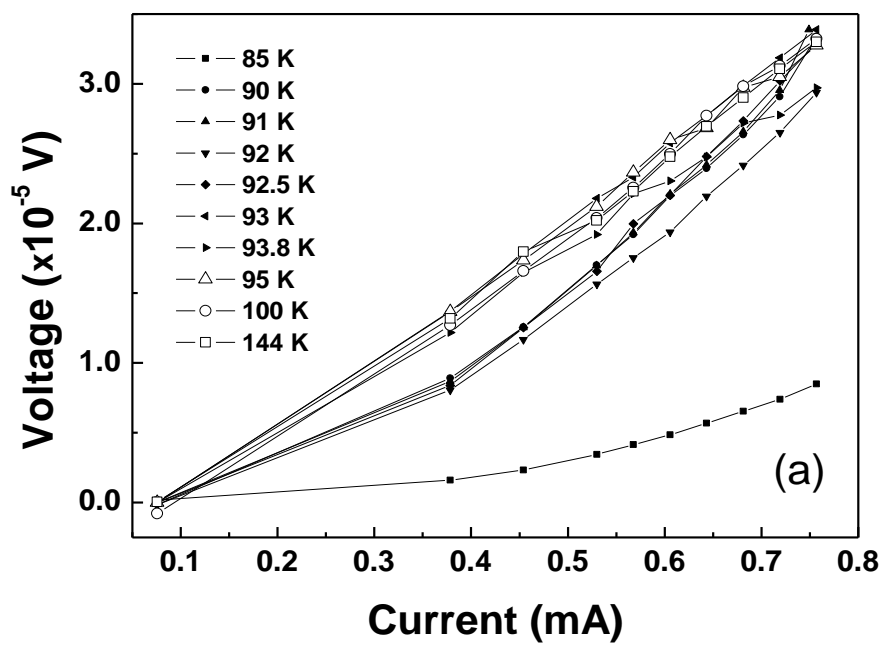

(a)

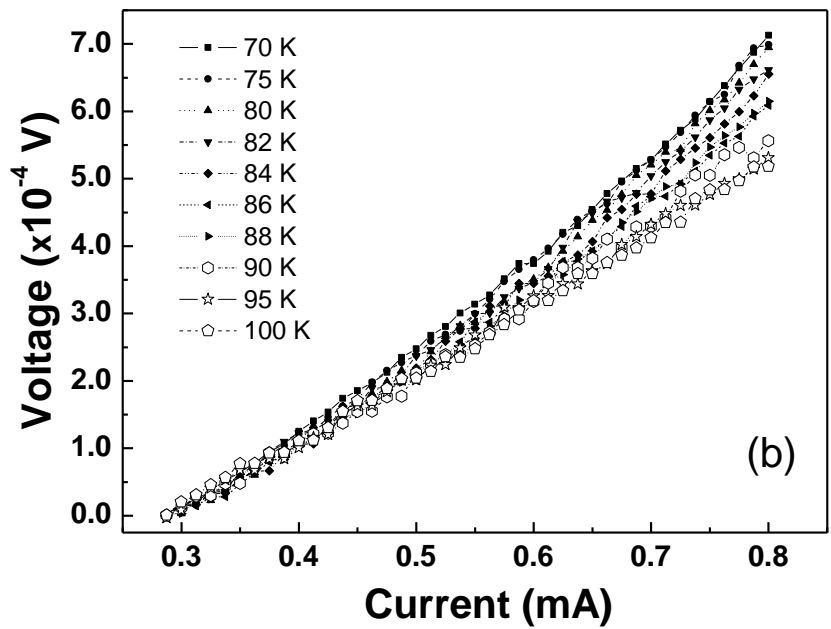

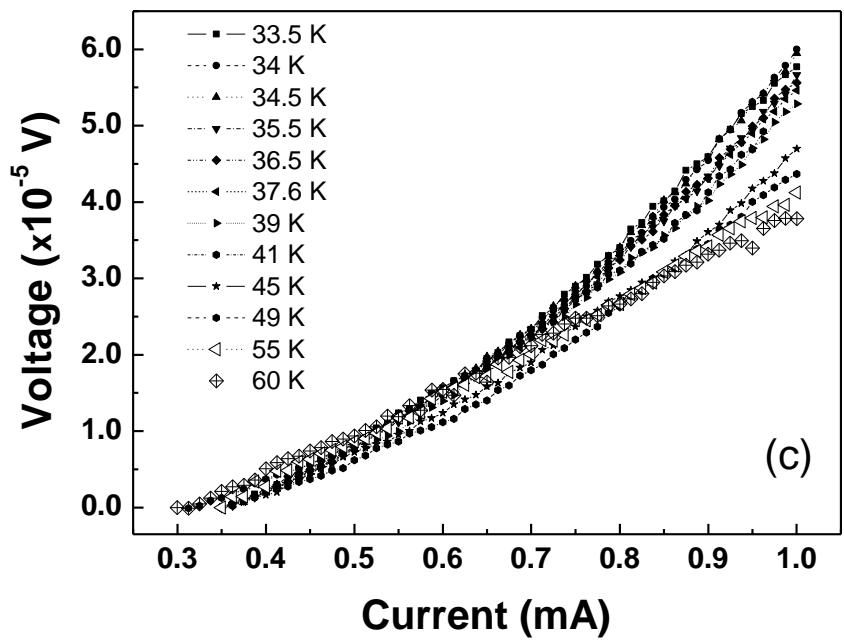

**Fig.2:** Current-voltage characteristics for three samples: $YBa_2Cu_3O_{7-\delta}$ (a), $Pr_{0.1}Y_{0.9}Ba_2Cu_3O_{7-\delta}$ (b), and $Pr_{0.3}Y_{0.7}Ba_2Cu_3O_{7-\delta}$ (c) taken at various temperatures.

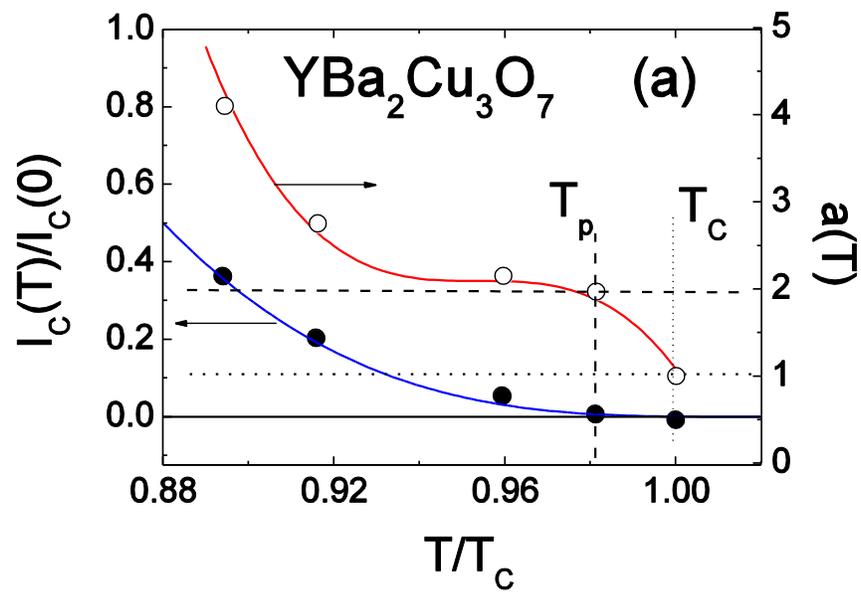

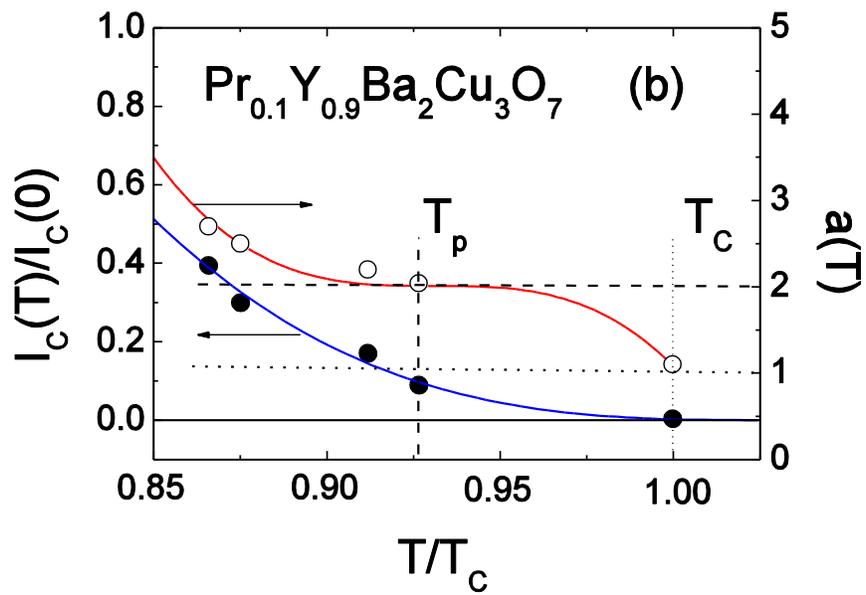

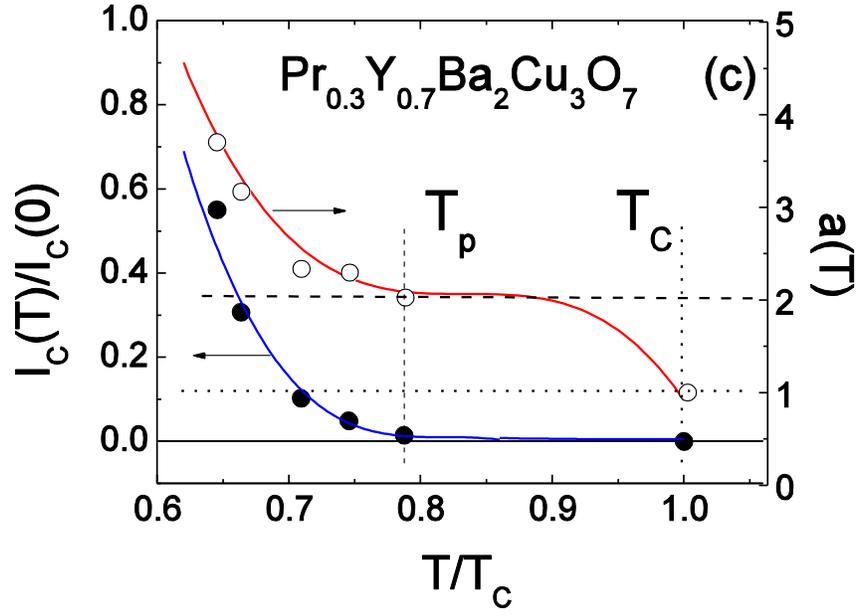

**Fig.3:** Temperature dependence of the experimental points (filled and open dots) along with theoretical fits (solid lines) for the critical current $I_C(T)$ and the power law exponent $a(T)$ deduced from the I×V data (see Fig.2) using Eqs.(1)-(4) for $YBa_2Cu_3O_{7-\delta}$ (a), $Pr_{0.1}Y_{0.7}Ba_2Cu_3O_{7-\delta}$ (b), and $Pr_{0.3}Y_{0.7}Ba_2Cu_3O_{7-\delta}$ (c). The horizontal (vertical) dotted and dashed lines show two transitions at $a(T_C)=1$ and $a(T_p)=2$ (which occur at corresponding temperatures $T_C$ and $T_p$).

**Table 1.** The values of the superconducting temperatures $T_C$ (extracted from our resistivity data) along with the fitting parameters ($T_p$ and $I_C(0)$) and deduced estimates ($\lambda_{ab}(0)$ and $\xi_{eff}(0)$) used to describe the temperature dependence of nonlinear current-voltage curves for three polycrystalline samples.

|  | $YBa_2Cu_3O_{7-\delta}$ | $Pr_{0.1}Y_{0.9}Ba_2Cu_3O_{7-\delta}$ | $Pr_{0.3}Y_{0.7}Ba_2Cu_3O_{7-\delta}$ |
|---|---|---|---|
| $T_C$, K | 92.9 | 89.5 | 53.9 |
| $T_p$, K | 91.2 | 82.7 | 43.1 |
| $I_C(0)$, mA | 0.6 | 0.4 | 0.2 |
| $\lambda_{ab}(0)$, µm | 0.14 | 0.15 | 0.17 |
| $\xi_{eff}(0)$, nm | 0.9 | 1.1 | 2.2 |